\documentclass[aps,prb,groupedaddress]{revtex4-1}
\usepackage{subfigure} 
\usepackage{amsmath}
\usepackage{graphics}
\usepackage{graphicx}
\usepackage{epsfig}
\usepackage{amssymb}
\usepackage{amsfonts}%
\usepackage{color}
\usepackage{url}
\providecommand{\U}[1]{\protect\rule{.1in}{.1in}}

\begin{document}

\title{Topological states in normal and superconducting $p$-wave chains}

\author{Mucio A. Continentino} 
\affiliation{Centro Brasileiro de Pesquisas F\'{i}sicas, Rua Dr. Xavier Sigaud, 150, Urca \\
22290-180, Rio de Janeiro, RJ, Brazil}
\author{Heron Caldas }
\affiliation{Departamento de Ci\^{e}ncias Naturais, Universidade Federal de S\~ao Jo\~ao Del Rei, \\ 36301-000, S\~ao Jo\~ao Del Rei, MG, Brazil}
\author{David Nozadze}
\author{Nandini Trivedi}
\affiliation{Department of Physics, The Ohio State University, Columbus, OH 43210, USA}


\begin{abstract}
We study a two-band model of fermions in a 1d chain with an antisymmetric hybridization that breaks inversion symmetry. 
We find that for certain values of its parameters, the $sp$-chain maps formally into a  $p$-wave superconducting chain, 
the archetypical 1d system exhibiting Majorana fermions. The eigenspectra, including the existence of zero energy modes 
in the topological phase, agree for both models. The end states too share several 
similarities in both models, such as the behavior of the localization length, the non-trivial topological index and robustness to disorder. 
However, we show by mapping the $s$- and $p$- fermions to two copies of Majoranas, that the excitations 
in the ends of a finite $sp$ chain are indeed conventional fermions though endowed with protected topological properties.
Our results are obtained by a scattering approach in a semi-infinite chain with an edge defect treated within the $T$-matrix approximation. We augment the 
analytical results with exact numerical diagonalization that allow us to extend 
our results to arbitrary parameters and also to disordered systems.
\end{abstract}

\maketitle


\section{Introduction}

In the search for a hardware for implementing quantum computation, Majorana fermions~\cite{majorana} have emerged as ideal candidates.
These quasi-particles appear in pairs and in case they are sufficiently separated they should be robust to quantum decoherence, 
a major obstacle for achieving a quantum computer~\cite{quantum}.
Majorana fermions have been shown to exist in the ends of $p$-wave superconducting chains~\cite{kitaev}. A model proposed by Kitaev
has played an important role in clarifying the nature of these quasi-particles and how they can appear in condensed matter 
systems~\cite{kitaev}. Majorana fermions are zero energy solutions of the Dirac equation with the special property that they 
are their own anti-particles~\cite{shen,bernevig}. The excitations at the ends of a $p$-wave superconducting chain depend 
on the quantum state of the system. This is determined by the ratio $|\mu|/2t$ between the chemical potential $\mu$ and 
the electron hopping $t$. If $|\mu|/2t<1$ the chain is superconducting with non-trivial topological properties. 
In this {\it weak pairing} phase it presents Majorana fermions at its ends. Otherwise, if $|\mu|/2t>1$, the chain is in 
the  {\it strong coupling} superconducting phase with trivial topological properties and has no end states~\cite{alicea,cdc}. 
This result can be obtained using either  Bogoliubov-de Gennes or Dirac equations with a space dependent order parameter~\cite{shen}, 
or more directly using a Majorana representation~\cite{kitaev}. 

In this paper, we study a non-interacting two-band model of a chain of atoms with electronic orbitals of angular momentum 
$l$ and $l+1$. Besides the hopping of electrons in the same orbital, we consider hybridization between different orbitals 
in neighboring sites. Since these orbitals have angular momentum differing by an odd number, their wave-functions have 
opposite parities. Then,  the hybridization between the orbitals in neighboring sites $i$ and $j$ is  antisymmetric, 
i.e, $V_{ij}=-V_{ji}$ or in $k$-space $V(-k)=-V(k)$. For simplicity, we consider a $sp$-chain for which the importance 
of the mixing of $sp$ bands for topological insulators has already been pointed out in different contexts, including that 
of the spin quantum Hall effect~\cite{zhang} and cold atoms~\cite{liu}.  The interest in these chains goes back to Schokley's 
work on surface states~\cite{shockley}. These states were investigated in detail~\cite{foo}, but until recently it lacked 
the theoretical tools for their understanding, specially
of the zero energy surface states~\cite{kane}.

A cold atom version of the $sp$-chain, the $sp$-ladder, has recently been investigated in light of the new concepts of 
topological insulators~\cite{liu}. Also a one dimensional Kondo lattice has been considered~\cite{coleman}. 
Both works consider interactions among the quasi-particles besides the topological aspects of the problem.

In this work, we concentrate on the simplest, non-interacting condensed matter system, a chain with two atomic 
orbitals per site with an antisymmetric hybridization between these orbitals.  We show that by tuning the parameters of this chain it is driven,
through a quantum phase transition~\cite{livro}, from a trivial  to a topological 
insulator~\cite{kane}. The nature of the end states in  semi-infinite $sp$-chains, is studied using a scattering 
approach that considers the end of the chain as a defect in a perfect crystal~\cite{foo}. The  scattering problem 
due to the defect has an associated $T$-matrix whose poles give the energies of the end states.

In spite of the different symmetries of the $sp$ and $p$-wave chains, we find a mapping between these problems.  
This mapping is purely formal as the two systems have different symmetries and physical properties. 
The former is a non-interacting band insulator and the latter a $p$-wave superconductor. This mapping in the case of semi infinite chains leads to interesting results regarding the nature of the end states. We show that 
both the semi-infinite $p$-wave  and  $sp$-chains  have zero energy end modes when these systems are in 
non-trivial topological phases. For the $p$-wave chain, this zero energy mode is immediately recognized as 
a Majorana fermion. For the $sp$-chain, the zero energy state share many properties with those of a Majorana mode. 
For example, the penetration length of the zero energy end modes of both models diverge at the topological 
transition separating the trivial from the non-trivial topological phases. Exact diagonalization results 
confirm the correlation length exponent obtained analytically and are able to extend the results to parameter 
values away from the specific symmetric point and also include the effects of disorder. 

Our paper is organized as follows.  In Sec. II, we  show that the eigenspectrum of the $sp$-chain with odd parity hybridization is 
identical to that of the $p$-wave superconducting chain for a specific choice of parameters. 
In Sec. III, using a $T$-matrix scattering approach we calculate the decay length of the end mode 
and compare with exact diagonalization results. While the analytical method is limited to a very 
specific choice of the parameters, the numerical method can also be extended to arbitrary values of these parameters and gives some insights on the nature of the
end states in the topological phase. In Sec. IV, we study robustness of the end states against disorder
using exact diagonalization approach. We find that the end states survive for any values of the strength of
the symmetric disorder. In Sec. V, we show that there are two unpaired Majorana fermions in the each end of the
$sp$-chain. These Majorana fermions can combine to form a conventional fermion with hybrid $sp$ character.
We also show that if the ends of the $sp$-chain  are connected o a bulk metallic reservoir with $s$-states only, one $p$-type Majorana appears on each end of the chain.     Finally, we conclude in Sec. VI.

\section{The $sp$ and the $p$-wave superconducting chains}

\begin{figure}[th]
\centering{\includegraphics[scale=0.3]{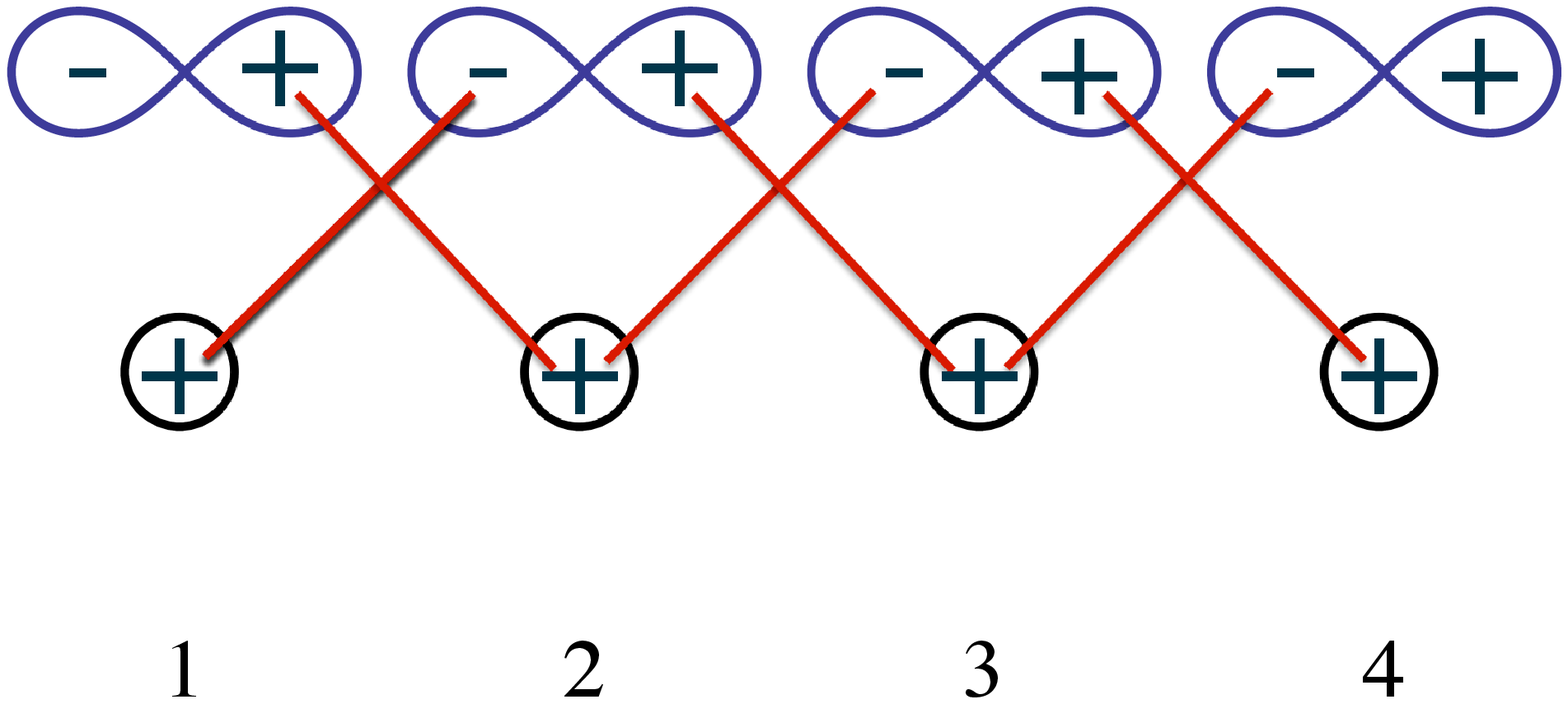}}
\caption{(Color
online) The $sp$-chain with four sites showing the parity of the orbitals and their hybridization. }%
\label{fig1}%
\end{figure}

Let us consider a one dimensional atomic system with two orbitals per site of angular momenta $l$ and $l+1$. 
Due to their different parities the hybridization between these orbitals in neighboring sites is antisymmetric. 
This system can be described by a two-band Hamiltonian with an antisymmetric hybridization. For simplicity, 
we imagine an $sp$-chain with the Hamiltonian given by~\cite{foo},
\begin{eqnarray}
\label{sp}
\mathcal{H}_{sp}&=& \epsilon^0_s \sum_j c^{\dagger}_j c_j + \epsilon^0_p \sum_j p^{\dagger}_j p_j- 
\sum_j t_s (c^{\dagger}_j c_{j+1} + c^{\dagger}_{j+1} c_{j})+ \sum_j t_p (p^{\dagger}_j p_{j+1} + p^{\dagger}_{j+1} p_{j}) \nonumber \\
&+&V_{sp} \sum_j (c^{\dagger}_j p_{j+1} -c^{\dagger}_{j+1} p_{j})) - V_{ps} \sum_j (p^{\dagger}_j c_{j+1} - p^{\dagger}_{j+1} c_{j})  
\end{eqnarray}
where $\epsilon^0_{s,p}$ are the centers of the $s$ and $p$ bands respectively. The $t_{s,p}$ represent 
the hopping of electrons to neighboring sites in the same orbital. $V_{sp}=V_{ps}$ and the antisymmetric 
nature of the hybridization between the $s$ and $p$ states in neighboring sites is taken into account 
explicitly in the Hamiltonian (see Fig.~\ref{fig1}). Due to the different parities of the orbital states, 
this hybridization is odd-parity, such that, $V_{sp}(-x)=-V_{sp}(x)$ or in momentum space $V_{sp}(-k)=-V_{sp}(k)$. 
Thus the mixing term breaks parity symmetry in spite of the fact that the chain is centro-symmetric. We take here the chemical potential of the $sp$-chain $\mu_{sp}=0$. 

One of our aims in this paper is to compare the $sp$-chain with Kitaev's model, so for completeness, 
we give below the Hamiltonian of the $p$-wave one-dimensional supercondutor~\cite{kitaev}. 
This model has been extensively studied and is the archetypical example of a system exhibiting Majorana fermions. 
It's Hamiltonian is given by~\cite{kitaev},
\begin{equation}
\label{kita}
\mathcal{H}=- \mu \sum_j c^{\dagger}_j c_{j} - t \sum_{j} ( c^{\dagger}_j c_{j+1}+ c^{\dagger}_{j+1} c_{j})  + \sum_{j} \Delta \left(    c_j c_{j+1} -      c_{j+1} c_{j} \right) +    \sum_{j}   \Delta^{*} \left(c^{\dagger}_{j+1} c^{\dagger}_{j} -c^{\dagger}_{j} c^{\dagger}_{j+1} \right) 
\end{equation}
where $t$ is a nearest neighbor hopping, $\mu$ the chemical potential. The anomalous nearest neighbor 
correlation function  $\Delta_{ij}=\Delta$ ($\Delta_{ji}=-\Delta$) is a pairing correlation that arises 
due to an  effective  antisymmetric attractive interaction between fermions in neighboring sites. 
The operators $c_i$ and $c_i^{\dagger}$ destroy and create fermions on site $i$ of the chain, respectively. 

At first sight the two Hamiltonians above have nothing in common since they represent two completely 
different problems. In particular the Kitaev superconducting model, Eq.~\ref{kita} has a broken U(1) symmetry, 
which is not the case of the $sp$-chain.  The hybridized two-band system described by Hamiltonian Eq.~\ref{sp} is a non-interacting problem, 
while that given by Eq.~\ref{kita} has a condensate ground state of electron pairs due to the attractive interaction. 
The only apparent similarity of the two problems is the presence of a term which breaks the parity symmetry, 
the odd-parity hybridization and the $p$-wave anomalous correlation function in Eqs.~\ref{sp} and~\ref{kita}, respectively. 
As we show below this feature alone will give rise  to important similarities in the electronic states of these systems.

Both Hamiltonians above can be easily solved to yield the energies of the new quasi-particles. For chains with 
periodic boundary conditions, this is accomplished using the translation invariance of the system and a simple 
Fourier transformation. Here we introduce the Greenian operator~\cite{foo} defined by, $\mathbf{G}=(\mathbf{1}-\mathbf{\mathcal{H}})^{-1}$.

For the $sp$ model, this is given by,

\begin{eqnarray}
\mathbf{G}_{sp}(k, \omega)=\frac{1}{D_{sp}} 
\left( \begin{array}{cc} \omega-\epsilon^0_p -2t_p \cos ka & 2iV_{sp} \sin ka \\ -2iV_{sp} \sin ka &   \omega -\epsilon^0_s + 2 t_s \cos ka  \\ \end{array} \right) \\  \nonumber
\end{eqnarray}
where
\begin{equation}
\label{reldisp}
D_{sp}(\omega)=(\omega -\epsilon^0_s + 2 t_s \cos ka)(\omega-\epsilon^0_p -2t_p \cos ka) - 4 V_{sp}^2 \sin^2 ka.
\end{equation}

For   the $p$-wave superconducting chain described by the Hamiltonian, Eq.~\ref{kita}, the Greenian can be written as:

\begin{eqnarray}
\mathbf{G}_{pw}(k, \omega)=\frac{1}{D_{pw}} 
\left( \begin{array}{cc} \omega + \epsilon_k & 2 i \Delta_0 \sin ka \\ -2 i \Delta_0 \sin ka &   \omega - \epsilon_k   \\ \end{array} \right)  \\ \nonumber
\end{eqnarray}
where
\begin{equation}
D_{pw}(\omega)=(\omega -\epsilon_k)(\omega+\epsilon_k) -4 \Delta_0^2 \sin^2 ka
\end{equation}
with $\epsilon_k= 2 t \cos ka - \mu$ and  $\Delta_0=|\Delta|$. 

The Greenians above yield a description of the excitations  
of the infinite $sp$ and $p$-wave superconducting chains. The energies of the quasi-particles for each model  
are obtained from the poles of the Greenians, i.e., from the equations $D_{sp}(\omega)=0$ and $D_{pw}(\omega)=0$. 
We can use them to establish a mapping between  the two models. For a special choice of the parameters of the 
$sp$-chain $t_s=t_p=t$ and $\epsilon^0_s=-\epsilon^0_p=\mu$, we find a complete formal correspondence between the two problems.  
The role of the antisymmetric hybridization $V_{sp}$  in the $p$-wave model is played by the antisymmetric gap $\Delta$, 
such that $V_{sp}=\Delta$.  With these identifications the eigenspectra of the two models are formally equivalent.

\section{Localization length of end mode:}

A fundamental feature of the $p$-wave superconducting model is that it exhibits Majorana fermions at the ends of 
the chain~\cite{kitaev}. 
In this section, we investigate the nature of the end states  by two methods: (A) a $T$-matrix scattering 
approach, and (B) exact diagonalization approach.

\subsection {T-matrix scattering approach:}
In order to study the edge excitations in the $sp$-chain, we consider a semi-infinite chain. The idea is to 
consider the {\it end} of the chain as a defect and treat this problem using a scattering approach following 
Ref.~\cite{foo}. The real space Greens function $G_S$ for the edge of the semi-infinite chain is obtained 
from a Dyson equation,
$$
G_S=G_0+G_0VG_S
$$
where $G_0$ is the Greens function for the perfect crystal (infinite chain) in the site representation. 
The potential $V$ is such that it creates a defect by {\it cleaving} the crystal say, between the sites $-1$ 
and $0$ of the infinite chain. It is given by,
 $$
V(0,-1)= \left( \begin{array}{cc} t_s & -V_{sp} \\ V_{ps} & -t_p \\ \end{array} \right)  \\ \nonumber
 $$
Using that the elements of $G_s$ between the two halves of the chain are zero one gets the end Greens function as
$$
G_S(m,m)=G_0(0)+G_0(m) T G_0(m)
$$
where $m$ is a site index. The poles of the  matrix $T(\omega)$ give the energies $E_S$ of the end states. These are obtained from the equation,
\begin{equation}
1-t_s G_{ss}(E_S)+2 V_{sp} G_{sp}(E_S) + t_p G_{pp}(E_S)=0
\end{equation}
where $G_{\alpha \beta}(E_S)=\sum_k G_{\alpha \beta}(k,E_S)e^{ika}$ with $a$ the lattice spacing ($\alpha$ and $\beta$=$s$,$p$). 
For the  particle-hole {\it  symmetric} $sp$-chain, with $t_s=t_p=t$ and $\epsilon^0_s=-\epsilon^0_p=\epsilon$,
the  above equation can be solved   analytically for $t=\epsilon$ and numerically for general $\epsilon \neq t$,  yielding
\begin{equation}
\label{esp}
E_S= 0.
\end{equation}
for $\epsilon/2t < 1$, independent of the value of $V_{sp}$.
\begin{figure}[htb]
\centering
\subfigure[]{
\includegraphics[width=0.4\linewidth]{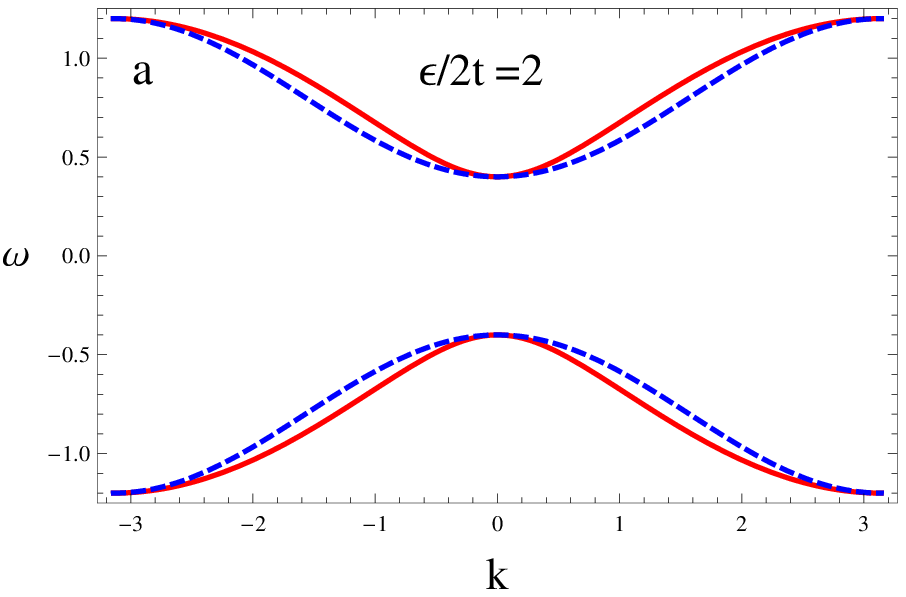}
\label{fig:subfig2a}
}\\
\subfigure[]{
\includegraphics[width=0.4\linewidth]{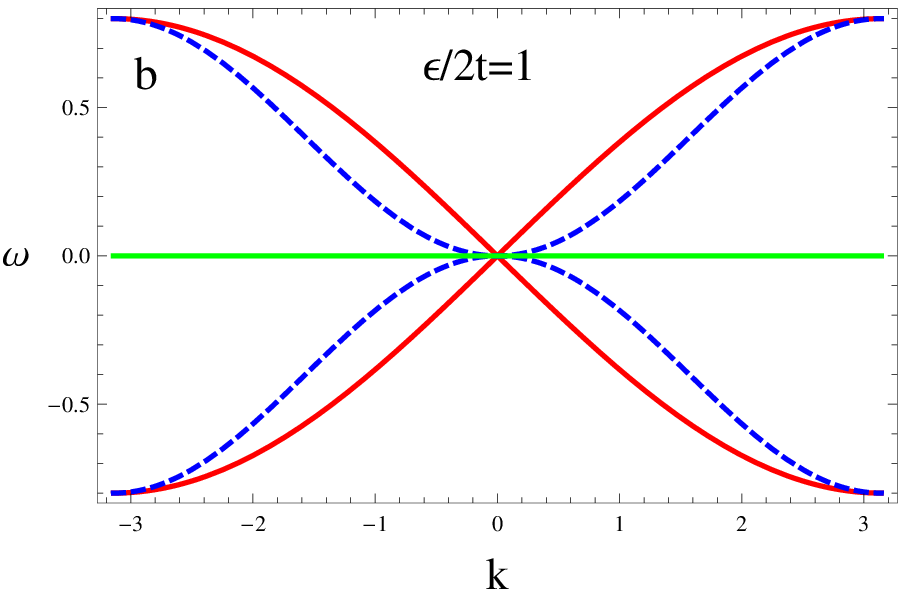}
\label{fig:subfig2b}
}\\ 
\subfigure[]{
\includegraphics[width=0.4\linewidth]{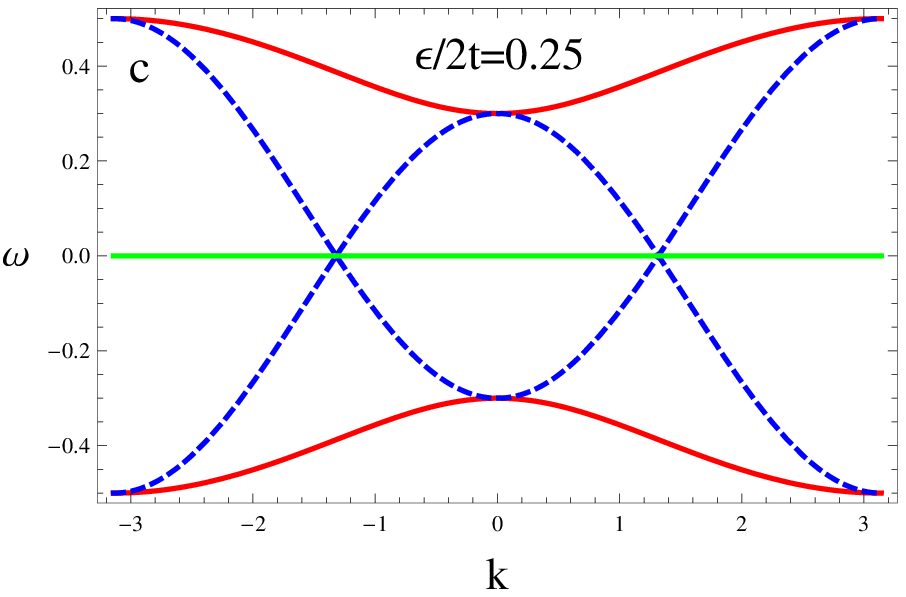}
\label{fig:subfig2c}
}
\caption{(Color
online) Dispersion relations of the 1d $sp$-model (roots of Eq.~\ref{reldisp}) for $V_{sp}=t$ and different 
values of the ratio $\epsilon/2t$. a) For $\epsilon/2t >1$, the system is a standard  band insulator with a gap 
between the valence and conduction bands. In this case the original bands in the absence of hybridization 
(dashed lines) do not overlap.  b) For $\epsilon/2t=1$ there is a topological transition between the trivial 
and non-trivial insulating states. The transition occurs with the closure of the gap between the bands. 
Near the center of the band, the excitations are Dirac-like fermions with a linear dispersion given by, 
$\omega =2 V_{sp} k$. c) For $\epsilon/2t<1$, the system is a topological insulator
with a zero energy mode on the end of the semi-infinite chain.}
\end{figure}
The decay of these end states,  assuming it is exponential inside the crystal, can be obtained using Schr\"odinger's equation~\cite{foo}.  
The characteristic penetration length $\xi$ for $V_{sp}=t$ is given by,
\begin{equation}
\label{xi}
\xi^{-1} \propto|\ln \frac{\epsilon}{2t}|.
\end{equation}
 The condition $\epsilon/2t<1$ constraints the argument of the logarithmic in Eq.~\ref{xi} being less 
than one and implies an overlap of the bands in the absence of hybridization (see Figs.~\ref{fig:subfig2a}).

The end states in the case of a generic $sp$-chain are the equivalent of the Shockley states in surface physics~\cite{shockley,beenakker}.
However, in the particle-hole symmetric case, i.e., for $t_s= t_p=t$ and $\epsilon^0_s=-\epsilon^0_p=\epsilon$ with $\epsilon/2t<1$, such that $E_S=0$, the zero energy modes correspond to the edge excitations of a topological phase of the $sp$-chain.  As we show in Sec. IV, for $\epsilon/2t=1$ the symmetric $sp$-chain has a quantum  phase transition from a topological insulator to a trivial insulating phase. The penetration length of the end mode  defined in Eq.~\ref{xi} diverges at this transition as,
\begin{equation}
\xi \propto  \frac{1}{(1-\frac{\epsilon}{2t})^{\nu}}
\label{xisp}
\end{equation}
with a critical exponent $\nu=1$.
For $\epsilon/2t<1$ (Fig.~\ref{fig:subfig2c}), where the original bands in the absence of hybridization overlap, 
the symmetric $sp$-chain is in a topologically non-trivial phase that corresponds to a topological insulator (see in Sec. IV). 
This phase has zero energy end modes, which disappear for $\epsilon/2t>1$, i.e., in the topologically trivial insulating phase (Fig.~\ref{fig:subfig2a}).
Now, using the mapping between the $sp$-chain and Kitaev's model  the results of the scattering approach 
for the former can be immediately transposed for the $p$-wave superconducting chain.
In the case of the $p$-wave model, the mapping implies $t_p=t_s=t$ and $\epsilon^0_s=-\epsilon^0_p=\mu$. 
This leads for the energy of the end mode,
\begin{equation}
\label{epw}
E_S=0,
\end{equation}
where the condition for the existence of this zero energy mode is $\mu/2t<1$. Comparing this result 
with  Kitaev's  solution of the $p$-wave model model, we notice that the $E_S=0$ mode can be recognized 
as the Majorana fermion at the end of the chain~\cite{kitaev}. Furthermore, the condition $|\mu|<2t$ refers to the region 
of the phase diagram corresponding to the topologically non-trivial weak pairing phase where zero energy modes 
are expected to exist.  Then, in the present approach,  the Majorana mode appears as an  end state arising 
from a scattering problem. 
As is known for Kitaev's model~\cite{kitaev}, and is obtained here for the $sp$-chain, the conditions for these systems to be in their topological non-trivial phases  with zero energy end modes do not  
involve $\Delta_0$ or $V_{sp}$, respectively, although these quantities must be finite.
Then,  in the case of Kitaev's model, the scattering approach that considers the end of a semi-infinite chain
as a scattering center leads to well known results for this model~\cite{kitaev}. 

As for the $sp$-chain, the exponential decay of the end state inside the superconductor has a characteristic penetration length $\xi$ given by,
 $\xi^{-1}=|\ln ( |\mu| /2t)|$ ($|\mu|<2t$). This  diverges at the weak-to-strong-pairing transition when $\mu=\mu_c=2t$ as,
\begin{equation}
\label{ximu}
\xi \propto \frac{1}{(\mu_c-\mu)^{\nu}}.
\end{equation}
with~\cite{exponent} $\nu=1$.

\subsection {Exact diagonalization approach:}

To verify the predictions of the  $T$-matrix approach  and to illustrate our results, we perform exact numerical diagonalization
of the Hamiltonian in Eq.~\ref{sp} using the transformation

\begin{align}
\label{tr1}
c_{j}&=\sum_{n}(u_{n}(j)\alpha_n+v_n(j)\beta_n)\nonumber\\ 
p_{j}&=\sum_{n}(-v_{n}(j)\alpha_n+u_n(j)\beta_n) \,,
\end{align}
where $\alpha$ and $\beta$ are the quasiparticle operators.
$u_n(j)$ and $v_n(j)$, which satisfy $u^2_n(j)+v^2_n(j)=1$ for
each site $j$, can be obtained from
\begin{align}
\label{bdg}
 \left( \begin{array}{cc}
h_1 & V_1  \\
V_2 & h_2  \end{array} \right)
\left(\begin{array}{c}
u_n(j)   \\
v_n(j)   \end{array}\right)=E_n
\left( \begin{array}{c}
u_n(j)   \\
v_n(j)   \end{array} \right)\,,
\end{align}
where $h_{1}u_n(j)=\epsilon^0_{s}u_n(j)-t_{s}(u_n(j+1)+u_n(j-1))$, $h_{2}v_n(j)=\epsilon^0_{p}v_n(j)+t_{p}(v_n(j+1)+v_n(j-1))$,
$V_{1}v_n(j)=-V_{sp}v_n(j-1)+V_{sp}v_n(j+1))$ and $V_{2}u_n(j)=V_{ps}u_n(j-1)-V_{ps}u_n(j+1))$. 
We numerically determine the eigenvalues $E_n$ and eigenvectors $(u_n(j),v_n(j))$ of Eq.~\ref{bdg} for the chain length  $L=256$.

\begin{figure}[th]
\centering{\includegraphics[scale=0.3]{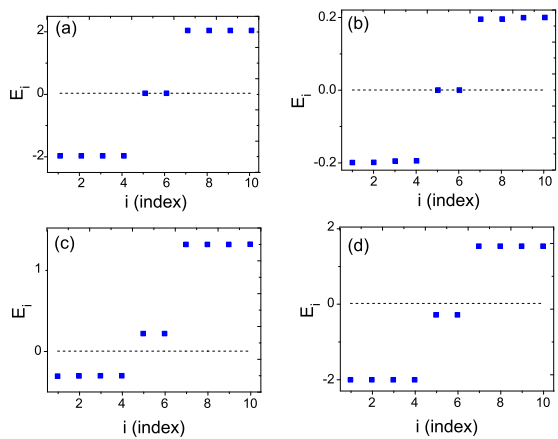}}
\caption{(Color
online) Eigenvalues versus  indices  for  different system parameters. (a)
 $t_s=t_p=V_{sp}=V_{ps}=1$  and $\epsilon _s^0=-\epsilon_p^0=0.5$, (b)
all parameters  as in (a) except $V_{sp}=V_{ps}=0.1$, (c) all parameters  as in (a) except  $t_p=0.4$,
and (d) all parameters  as in (a) except $\epsilon_p^0=0$. Number of modes is 512 and a limited number of them around zero is shown. }%
\label{fig3}%
\end{figure}

\begin{figure}[htb]
\centering
\subfigure[]{
\includegraphics[width=0.4\linewidth]{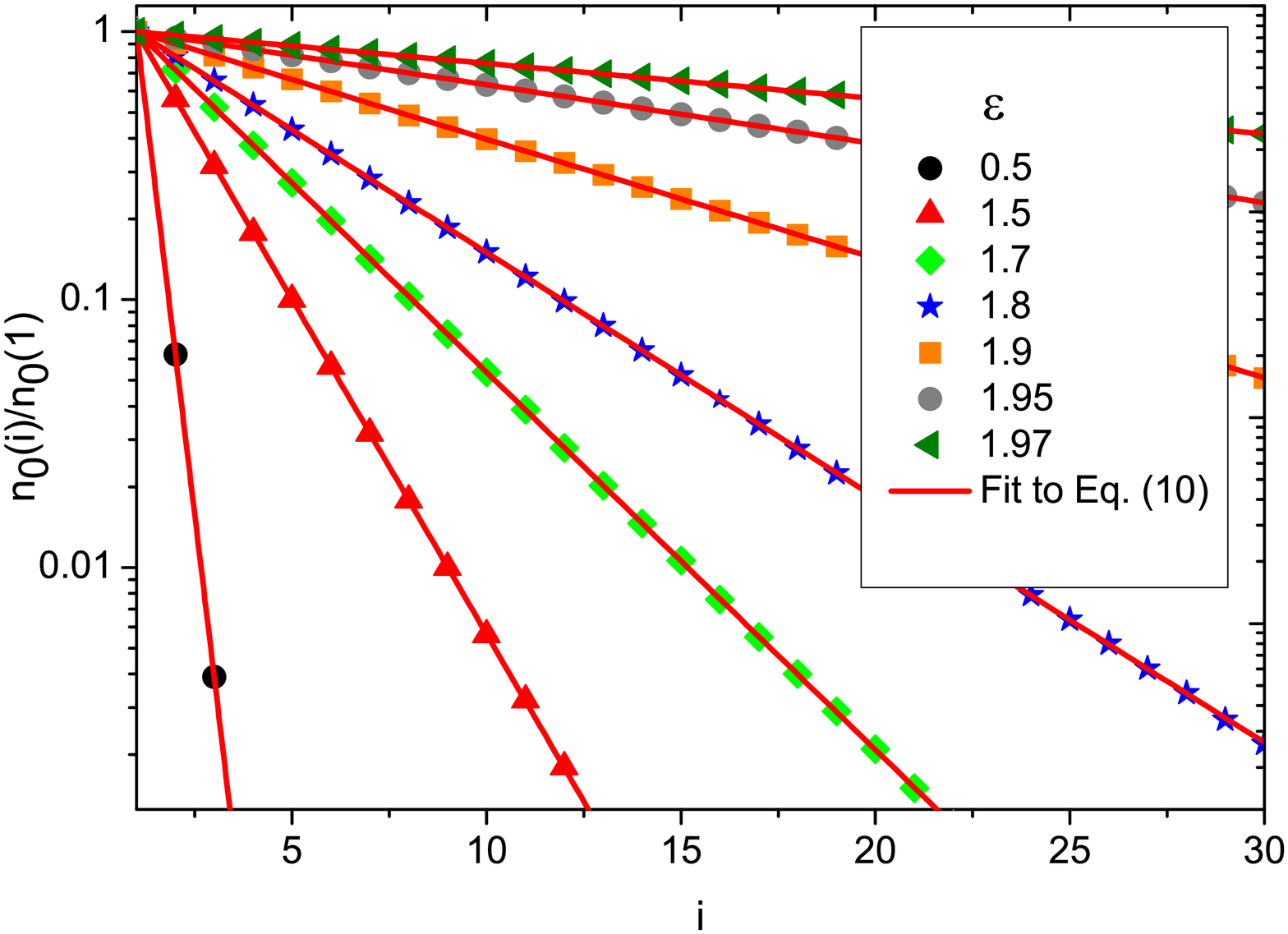}
\label{fig:subfig4a}
}
\subfigure[]{
\includegraphics[width=0.4\linewidth]{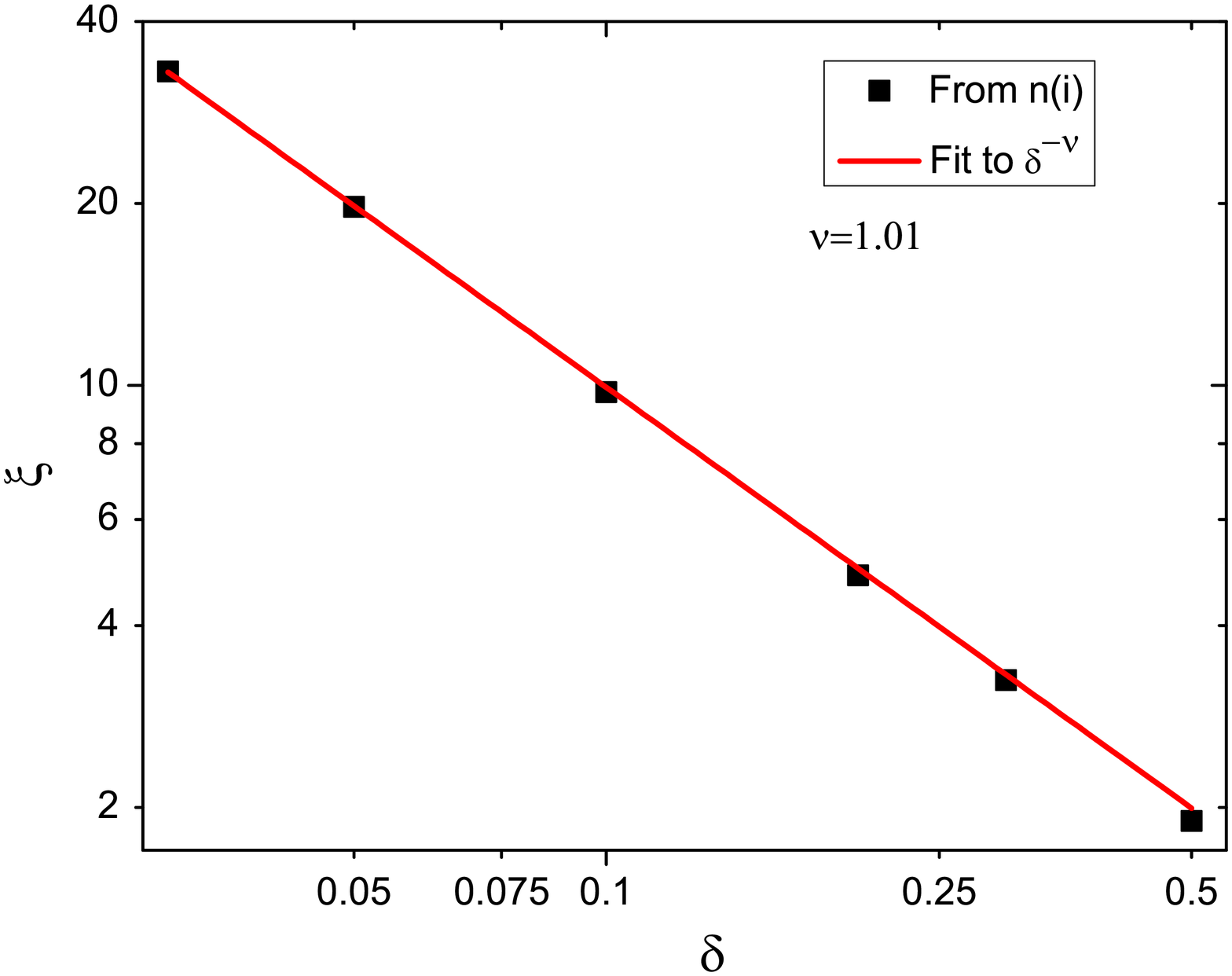}
\label{fig:subfig4b}
}
\caption{(Color
online) (a) The site dependence of the weights of  the zero modes in the
topological phase. The chain with  the length of $L=256$ is considered with the  system parameters: $t_s=t_p=V_{sp}=V_{ps}=1$ and for several values of $\epsilon=\epsilon _s^0=-\epsilon_p^0$. The fit by the solid straight lines imply that the quantity in the ordinate is proportional to $\exp(-l_i/\xi)$ where $l_i=i a$ is the position of the $i$th atom and $a$ is the lattice parameter.
(b) figure shows the values of $\xi$ obtained from these curves versus distance $\delta$ from quantum criticality. The solid line is a fit to a power law, resulting in the correlation length exponent $\nu= 1.01$).}%
\end{figure}

In Fig.~\ref{fig3}, we consider different choices for the parameters and 
calculate the eigenvalues $E_n$ and eigenvectors $(u_n(j),v_n(j))$ of Eq.~\ref{bdg} 
in the topological phase $(|\epsilon|<2t)$. As the result,
we obtain two zero eigenvalues as expected (see Figures~\ref{fig3}(a,b)). 
Notice that if the system is not in the symmetric case the modes move away from zero energy as shown in 
Figs~\ref{fig3}(c,d).

Next, we  consider $t_s=t_p=V_{sp}=V_{ps}=1$ and $\epsilon=\epsilon _s^0=-\epsilon_p^0$ and calculate the weight  of zero modes defined by $n_0(i)=\sum_{k=1}^2(u^2_k(i)+v^2_k(i))$
as a function of site number $i$ for different values of $\epsilon$ in the topological phase $(|\epsilon|<2t)$. Here, $u_{1,2}(i)$ and $v_{1,2}(i)$ are eigenvectors
corresponding to two zero energies. The resulting data is described 
by Eq.~\ref{xisp} very well (see Figure~\ref{fig:subfig4a}).  The penetration length $\xi$ can be determined
from fits to Eq.~\ref{xisp} (Fig.~\ref{fig:subfig4a}). Figure.~\ref{fig:subfig4b} shows how the penetration length $\xi$ changes 
with distance from criticality $\delta= 2t-\epsilon$. The data can be fitted to the power-law
$\xi\sim\ \delta^{-\nu}$. By fitting we extract the exponent $\nu=1.01$ which is in agreement 
with the prediction Eq.~\ref{xisp}.

\begin{figure}[th]
\centering{\includegraphics[scale=0.3]{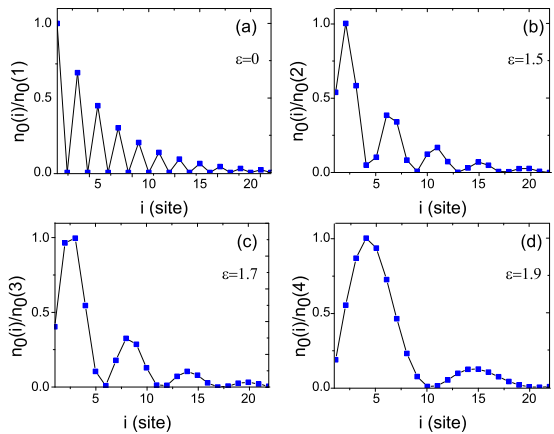}}
\caption{(Color
online) The site dependence of weights of  the zero modes  in the
topological phase. The zero mode weights   undergo damped oscillations. The period of oscillations increases as  approaching criticality. 
 The chain with  the length of $L=256$ is considered with parameters: $t_s=t_p=1$, $V_{sp}=V_{ps}=0.1$ and 
several values of $\epsilon=\epsilon _s^0=-\epsilon_p^0$.}%
\label{fig5}%
\end{figure}

We now solve  Eq.~\ref{bdg} for  $t_s=t_p=1$ and $V_{sp}=V_{ps}=0.1$ for several values of $\epsilon=\epsilon _s^0=-\epsilon_p^0$. 
In this case, the zero modes undergo damped oscillations (Fig.~\ref{fig5}) in contrast with the case considered previously where zero modes decay without 
oscillations (Fig.~\ref{fig:subfig4a}).   For $\epsilon=0$,   weights of  the zero modes are zero on even lattice sites  
and decay exponentially (Fig.~\ref{fig5}(a)) in agreement with the result given in 
Ref.~\cite{coleman}. This is in contrast with cases when $\epsilon \neq 0$ where zero modes decays in different manner. The zero modes on each even lattice sites
are not zero anymore. The period of the oscillations increases as approaching the critical point (Fig.~\ref{fig5} (b,c,d)).

\section {Robustness of the edge modes against disorder}

\begin{figure}[th]
\centering{\includegraphics[scale=0.3]{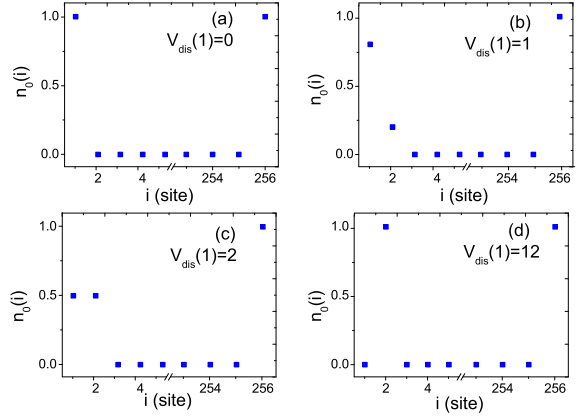}}
\caption{(Color
online) The figures show, going from (a) to (d),  how the weight of  the zero mode
shifts to the second site when the strength of disorder on the first site
increases.  The chain with  the length of $L=256$ is considered with  parameters:  $t_s=t_p=V_{sp}=V_{ps}=1$, $\epsilon _s^0=-\epsilon_p^0=0$
and several values of disorder strength $V_{{\rm{dis}}}(1)$ presented only on the first site
 $(\epsilon _s^0 \rightarrow \epsilon _s^0 +V_{{\rm{dis}}}(1)$ and  $\epsilon_p^0  \rightarrow  \epsilon _p^0 -V_{{\rm{dis}}}(1)))$.}%
\label{fig6}%
\end{figure}

In this section,  we explore the robustness of the edge modes against a diagonal disorder. In the absence of disorder zero modes are localized on
the ends of the chain (Fig. \ref{fig6}(a)). 

\begin{figure}[th]
\centering{\includegraphics[scale=0.3]{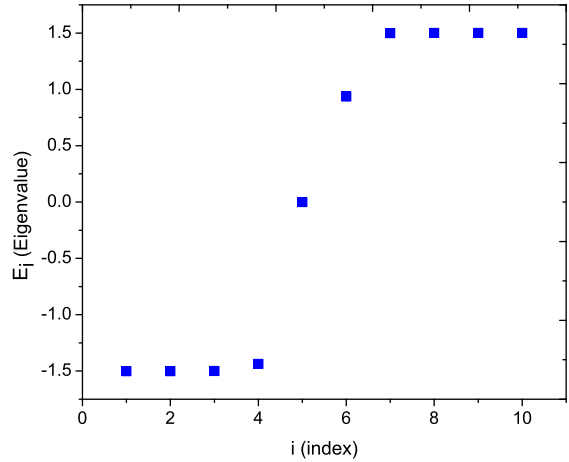}}
\caption{(Color
online) Eigenvalues versus  indices. Only one zero mode survives when disorder is 
presented non-symmetrically.  The following parameters are taken: 
$t_s=t_p=V_{sp}=V_{ps}=1$  and $\epsilon _s^0=-\epsilon_p^0=0.5$ in the presence of disorder
 $V_{{\rm{dis}}}(1)=1$ only on the first site
$(\epsilon _s^0 \rightarrow \epsilon _s^0 +V_{{\rm{dis}}}(1)$ and  $\epsilon_p^0  \rightarrow  \epsilon _p^0$).
 Number of modes is 512 and a limited number of them around zero is shown. }%
\label{fig7}%
\end{figure}

Let us now consider the effect of single symmetric disorder $V_{{\rm{dis}}}(1)$, which is present only
on the first site. The symmetric disorder means that disorder acts on the centers of the $s$ and $p$ bands in the 
following way:  $\epsilon _s^0 \rightarrow \epsilon _s^0 +V_{{\rm{dis}}}(1)$ 
and  $\epsilon_p^0  \rightarrow  \epsilon _p^0 -V_{{\rm{dis}}}(1)$. As the result, we find that both zero modes always
survive for any value of the disorder strength (Fig.~\ref{fig6}). The zero mode weight
is shared between the first and second sites for small disorder potential (Fig.~\ref{fig6}(b,c)), and is totally shifted to the second
site for large disorder strength (Fig.~\ref{fig6}(d)).
The zero mode localized on the last site is not affected  when disorder is
present on the first site (Fig.~\ref{fig6}).  When   non-symmetric disorder (e.g. $\epsilon _s^0 \rightarrow \epsilon _s^0 +V_{{\rm{dis}}}(1)$ 
and  $\epsilon_p^0  \rightarrow  \epsilon _p^0$) is present on the first site, only one zero mode localized on the last site
survives (Fig.~\ref{fig7}). 

\section{ End modes: Majoranas or conventional fermions?}

In the discussion below, we investigate the topological character of the elementary excitations.
The topological nature of the insulating phases of the $sp$-chain for the symmetric case $t_s= t_p=t$ and $\epsilon^0_s=-\epsilon^0_p=\epsilon$ can be established by calculating the winding number of these phases~\cite{shen,alicea}. 
We start writing the fermion operators of the $sp$-chain Hamiltonian in terms of Majorana fermions~\cite{kitaev,shen}.
These are given by the following definitions,
\begin{eqnarray}
\label{majo}
c_i=\frac{\alpha_{Bi} + i \alpha_{Ai}}{\sqrt{2}} \nonumber \\
p_i=\frac{\beta_{Bi} + i \beta_{Ai}}{\sqrt{2} }
\end{eqnarray}
and similar equations for their complex conjugates, noticing that the Majorana particles are their own antiparticles, i.e., $\alpha_{Ai}^{\dagger}=  \alpha_{Ai}$, $\alpha_{Bi}^{\dagger}=  \alpha_{Bi}$, $\beta_{Ai}^{\dagger}=  \beta_{Ai}$ and $\beta_{Bi}^{\dagger}=  \beta_{Bi}$. In the Majorana basis the Hamiltonian of the $sp$-chain can be written as (for $t_s= t_p=t$ and $\epsilon^0_s=-\epsilon^0_p=\epsilon$),
\begin{eqnarray}
H_{sp}&=&i \epsilon \sum_{i=1}^N (\alpha_{Bi} \alpha_{Ai} - \beta_{Bi} \beta{Ai} )-it \sum_{i=1}^{N-1} (\alpha_{Bi} \alpha_{Ai+i} - \alpha_{Ai} \alpha_{Bi+1} )
+it \sum_{i=1}^{N-1} (\beta_{Bi} \beta_{Ai+i} - \beta_{Ai} \beta_{Bi+1} ) \nonumber \\
&+&iV \sum_{i=1}^{N-1} (\alpha_{Bi} \beta_{Ai+1}-\beta_{Bi}\alpha_{Ai+1}-\alpha_{Ai}\beta_{Bi+1}+\beta_{Ai}\alpha_{Bi+1}).
\end{eqnarray}
Next we apply a Fourier transformation. For this purpose we introduce,
\begin{eqnarray}
\alpha_{Bi}=\sum_k \alpha_k^B e^{ikx_i}
\end{eqnarray}
and similarly for the other Majoranas. Using the properties of the Majorana operators we see that $\alpha_{-k}^B=\alpha_{k}^{B \dagger}$. On the other hand,  the $k=0$ mode is a Majorana since,
$\alpha_{0}^B=\alpha_{0}^{B \dagger}$.
We finally get,
\begin{eqnarray}
H_{sp}= i \left[ \sum_k E_k \alpha^B_k\alpha^A_{-k} - \sum_k E_k \beta^B_k \beta^A_{-k}-2i \sum_k V_{sp}(k)\beta^A_k \alpha^B_{-k}+2i \sum_k V_{sp}(k)\beta_k^B \alpha^A_{-k}\right]
\end{eqnarray}
where $E_k=\epsilon -2t \cos ka$ and $V_{sp}(k)=V_{sp} \sin ka$. This Hamiltonian can be written in matrix form in the following way, $H_{sp}=\Psi H \Psi^{\dagger}$, where

\[H(k)=
\left(
\begin{array}{ccccc}
  0   &  -E_k & 0 & -2i V_{sp}(k) \\
 E_k & 0 & 2i V_{sp}(k) & 0 \\
 0 & -2iV_{sp}(k) & 0 & E_k \\
 2i V_{sp}(k) & 0 & - E_k & 0
\end{array}
\right)
\]
and $\Psi= (\alpha^A_k, \alpha^B_k, \beta^A_k, \beta^B_k)$. This Hamiltonian  belongs to the class AIII of chiral Hamiltonians~\cite{aiii}. The winding number $\mathcal{M}$ is defined as the product of the signs of the Pfaffians of the matrix $H(k)$ at $k=0$ and $k=\pi/a$. We have, $Pf[H(k=0)]=\epsilon-2t$ and $Pf[H(k=\pi/a)]=\epsilon+2t$, such that $\mathcal{M}=$sgn$(\epsilon-2t)$sgn$(\epsilon+2t)$. The trivial non-topological phase is characterized by $\mathcal{M}=+1$ while the topological phase is characterized~\cite{shen,alicea} by $\mathcal{M}=-1$. Then, in the case $t_s= t_p=t$ and $\epsilon^0_s=-\epsilon^0_p=\epsilon$ and for $\epsilon<2t$ the $sp$-chain is a topological insulator and for $\epsilon=2t$ there is a quantum topological transition from this state to a trivial insulating one with no special (topological)  properties, as shown in~Fig.\ref{fig:subfig2b}. Notice that, if we take $t_s= t_p=t$ and $\epsilon^0_s=-\epsilon^0_p=\mu$, with $t$ and $\mu$ referring to parameters of Kitaev's $p$-wave model, the results above translate directly into known results for this model~\cite{alicea}.  As can be seen in the derivation above, the criterion for the system to be in the topological phase does not involve directly $V_{sp}$.

In order to obtain the end modes of the chain, we introduce two new operators~\cite{coleman}
\begin{eqnarray}
l_i=c_i+p_i  \nonumber \\
r_i=c_i-p_i.
\end{eqnarray}
Using Eqs.~\ref{majo} these can be rewritten in terms of Majorana operators as,
\begin{eqnarray}
l_i=\gamma^{+}_{Bi}+i \gamma^{+}_{Ai}  \nonumber \\
r_i=\gamma^{-}_{Bi}+i \gamma^{-}_{Ai}
\end{eqnarray}
where we introduced new composite or hybrid $sp$ Majorana operators given by,
\begin{eqnarray}
\gamma^{\pm}_{A/Bi}= \alpha_{A/Bi} \pm \beta_{A/Bi}.  
\end{eqnarray}
In terms of these hybrid $sp$ Majorana operators, the symmetric ($t_s=t_p=t$,  $\epsilon_0^s = -\epsilon_0^p=\epsilon$)  $sp$-chain Hamiltonian can be written as,
\begin{eqnarray}
\mathcal{H}_{sp}&=& i\frac{\epsilon}{2}\sum_{i=1}^{N}(\gamma^{+}_{Bi} \gamma^{-}_{Ai}-
\gamma^{+}_{Ai} \gamma^{-}_{Bi}) +
2i(V-t)\sum_{i=1}^{N-1}(\gamma^{-}_{Bi}\gamma^{+}_{Ai+1}-\gamma^{-}_{Ai}\gamma^{+}_{Bi+1})\nonumber \\
&&-2i(V+t)\sum_{i=1}^{N-1}(\gamma^{+}_{Bi}\gamma^{-}_{Ai+1}-\gamma^{+}_{Ai}\gamma^{-}_{Bi+1}).
\end{eqnarray}
Let us consider the point in the phase diagram ($\epsilon=0$, $V=t$) inside the topological insulating phase. The Hamiltonian is given by,
\begin{eqnarray}
\mathcal{H}_{sp}= -4it\sum_{i=1}^{N-1}(\gamma^{+}_{Bi}\gamma^{-}_{Ai+1}-\gamma^{+}_{Ai}\gamma^{-}_{Bi+1}).
\end{eqnarray}
Notice that the Majorana operators $\gamma^{-}_{B1}$ and $\gamma^{-}_{A1}$ do not enter the Hamiltonian 
so that there are two unpaired hybrid $sp$ Majorana fermions in the end of the chain. The same occurs for 
$\gamma^{+}_{BN}$ and $\gamma^{+}_{AN}$ on the opposite end of the chain. However, two Majoranas  in the 
same extremity can combine to form a single conventional fermionic quasi-particle with hybrid $sp$ character. On the 
left end, we have $l_1= \gamma^{-}_{B1}+i \gamma^{-}_{A1}=c_1-p_1$.
On the right end, we have, $r_N=\gamma^{+}_{BN}+ i \gamma^{+}_{AN}=c_N+p_N$, a total of two electronic 
states one in each end of the chain. These zero energy modes are the protected modes of the topological 
insulator phase of the symmetric $sp$-chain as shown in previous sections.

Now imagine the ends of this finite $sp$-chain are connected to a bulk metallic reservoir with $s$-states only. Referring to the $s$-electrons of the reservoir  as $b$, the Hamiltonian which couples both systems can be written as~\cite{hibri,vernek},
 \begin{equation}
 H_{cb}= v(l^{\dagger}b+b^{\dagger}l) +v(r^{\dagger}b+b^{\dagger}r),
\end{equation} 
where $v$ is the tunneling element connecting the electrons of the reservoir  and those, one at each end, of the chain.
Substituting the expressions above for  {\it left} $l$ and {\it right} $r$ electrons, $l=c-p$ and $r=c+p$ in this equation, we obtain,
 \begin{equation}
 H_{cb}= 2v(c^{\dagger}b+b^{\dagger}c). 
 \end{equation} 
Then one electron $s$, half from each end of the chain tunnels to the reservoir leaving two $p$-type Majoranas, one at each end of the chain.

\section{Conclusions}

We have shown that the infinite, spinless, symmetric $sp$-chain maps formally in a model for  
a $p$-wave superconducting chain. In the region of parameter space where this mapping holds, 
the $sp$-chain presents a quantum phase transition (QPT)  from a topological insulator to a 
topologically trivial insulating phase. This QPT occurs when  the ratio  between the center 
of the bands and the hopping integrals, $\epsilon/2t=1$.  At the QFT, near the $\Gamma$ 
point ($k=0$), the spectrum is  linear with Dirac-like excitations. In the topological 
insulating phase, in the end of $sp$-chain there are two unpaired hybrid $sp$ Majorana fermions 
which can combine to form a single conventional fermion. This fermion mode shares many properties with the 
Majorana modes at the ends of a $p$-wave superconducting chain such as the behavior of the localization length,
the non-trivial topological index and robustness to disorder. 
We also show that in the topological insulator phase, if the ends of the $sp$-chain  are connected to a metallic reservoir with $s$-states only,  $p$-type Majoranas appear on each end of the chain.

Generically, for values of the parameters of the $sp$-chain 
where the mapping with Kitaev's does not hold, the $sp$-chain may still have end modes not 
necessarily of zero energy. In this case these modes can be identified with Schokley 
states~\cite{shockley},  as studied in surface physics~\cite{shockley,beenakker,coleman}.

The  important feature of both $sp$-chain and $p$-wave models that give rise to a common 
behavior is an antisymmetric term in their Hamiltonian that breaks inversion symmetry in
an otherwise centro-symmetric lattice.

We have shown that the scattering approach that treats the end of a semi-infinite chain 
as a defect in an infinite chain turns out to be useful  for studying end modes. 
It can be extended to higher dimensions~\cite{foo} and represents an additional tool  
to study these modes besides the usual BdG~\cite{read} or the Jackiw-Rebbi methods~\cite{jak} 
used in conjunction with Dirac equation~\cite{shen}.

 {\acknowledgments   We wish to thank the Brazilian agencies, CAPES, FAPERJ, CNPq and FAPEMIG for financial support. We would like to thank Jason Alicea and Alexander Altland for useful discussions. HC acknowledges the hospitality of CBPF where this work was done.
We (DN and NT) would like to acknowledge funding from Grant No. NSF  DMR-1309461.}

\end{document}